\def\etbr{$\kappa$-(BE\-DT\--TTF)$_2$\-Cu\-[N\-(CN)$_{2}$]Br}
\def\etcl{$\kappa$-(BE\-DT\--TTF)$_2$\-Cu\-[N\-(CN)$_{2}$]Cl}
\def\etbrcl{$\kappa$-(BE\-DT\--TTF)$_2$\-Cu\-[N\-(CN)$_{2}$]\-Br$_{x}$Cl$_{1-x}$}
\def\cm{cm$^{-1}$}
\def\Ueff{$U_{\rm eff}$}
\documentclass[prb,twocolumn,showpacs]{revtex4}
\usepackage[dvips]{graphicx}
\begin{document} 
\title{Bandwidth-controlled Mott transition in \etbrcl:\\
I. Optical studies of localized charge excitations}
\author{Daniel Faltermeier}
\author{Jakob Barz}
\author{Michael Dumm}
\author{Martin Dressel}
\email{dressel@pi1.physik.uni-stuttgart.de}
\affiliation{1.~Physikalisches Institut, Universit{\"a}t Stuttgart,
Pfaffenwaldring 57, 70550 Stuttgart Germany}
\author{Natalia Drichko}
\affiliation{1.~Physikalisches Institut, Universit{\"a}t Stuttgart,
Pfaffenwaldring 57, 70550 Stuttgart Germany} \affiliation{Ioffe
Physico-Technical Institute Russian Academy of Science
Politeknicheskaya 26, 194021 St.Petersburg, Russia}
\author{Boris Petrov}
\author{Victor Semkin}
\author{Rema Vlasova}
\affiliation{Ioffe Physico-Technical Institute Russian Academy of Science
Politeknicheskaya 26, 194021 St.Petersburg, Russia}
\author{C\'ecile Meziere}
\author{Patrick Batail}
\affiliation{Laboratoire CIMI, FRE 2447 CNRS-Universit{\'e} d'Angers,
B{\^a}t.\ K, UFR Sciences, 2 Bd.\ Lavoisier, 49045 Angers, France}
\date{\today}
\begin{abstract}
Infrared reflection measurements of the half-filled two-dimensional
organic conductors $\kappa$-(BEDT-TTF)$_2$Cu[N(CN)$_{2}$]Br$_{x}$Cl$_{1-x}$
were performed as a function of
temperature ($5~{\rm K}<T<300$~K) and Br-substitution ($x=0\%$, 40\%,
73\%, 85\%, and 90\%) in order to study the metal-insulator transition.
We can distinguish absorption processes due to itinerant and localized
charge carriers. The broad mid-infrared absorption has two
contributions: transitions between the two Hubbard bands and intradimer
excitations from the charges localized on the (BEDT-TTF)$_2$ dimer.
Since the latter couple to intramolecular vibrations of BEDT-TTF, the
analysis of both electronic and vibrational features provides a tool to
disentangle these contributions and to follow their temperature and
electronic-correlations dependence. Calculations based on the cluster
model support our interpretation.
\end{abstract}


\maketitle
%
%
%
%
\section{Introduction}
The physics of strongly correlated electron systems is a very active
field in solid-state science where the vicinity of Mott-insulating,
magnetically ordered, and superconducting ground states is most
intriguing. These effects are intensively studied for transition-metal
oxides, in particular high-temperature superconductors, and organic
conductors. It is extremely interesting that all of these materials
show a similar competition between ordered antiferromagnetic and
superconducting phases: this suggests common physics, while the
chemistry of the compounds and the origin of the conducting electrons
is different.\cite{Jerome94,Ishiguro98,McKenzie97} These facts have
initiated our investigation of the  molecular conductors of the
BEDT-TTF family as model compounds to study physics of correlated
electrons close to the Mott transition in two
dimensions.\cite{Seo04,Dressel04}
\begin{figure}
\centering
\includegraphics[width=85mm]{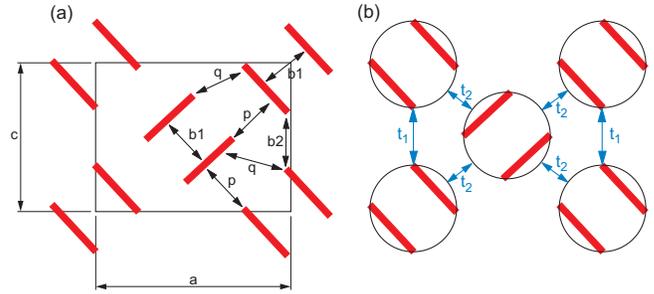}
\caption{\label{fig:kappapattern1}(a) Structural arrangement of
the BEDT-TTF molecules in the $\kappa$-phase (looking along the
molecular axis); the size of the unit cell is approximately
12.9~\AA\ in $a$-direction, 8.5~\AA\ along the $c$ axis, and
30.0~\AA\ in the third direction.  The overlap integrals are
labelled according to Mori {\it et al.}
(Ref.~\protect\onlinecite{Mori99}). The interdimer overlap
integral b1 is around 0.027, along the dimer chains ${\rm
b2}\approx0.010$, while p and q link orthogonal molecules with
approximately 0.011  and 0.004, respectively. (b) Triangular
lattice for the dimer model of $\kappa$-(BEDT-TTF)$_2X$. There is
hopping along the stacks $t_1$ and along the diagonals $t_2$.}
\end{figure}

\begin{figure}
\centering
\includegraphics[width=85mm]{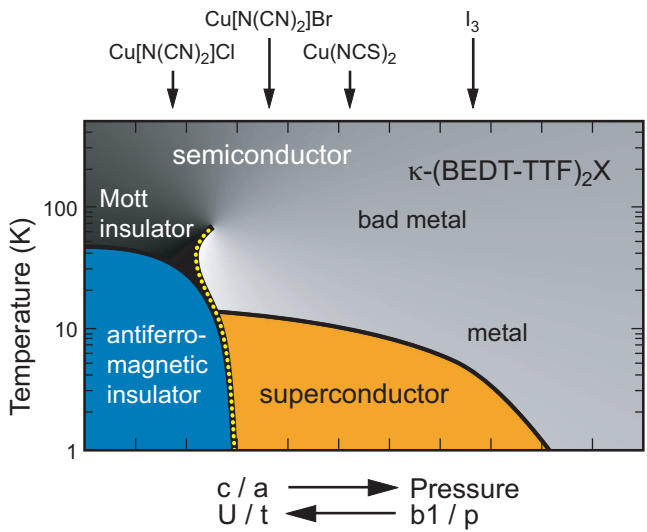}
\caption{\label{fig:ETphasediagram}Schematic phase diagram of the $\kappa$-phase
salts (BEDT-TTF)$_2X$. Instead of tuning the external pressure, the
same ambient-pressure ground state can be achieved by modifying the
anions $X$. The arrows indicate the approximate position of  \etcl,
 \etbr, $\kappa$-(BEDT-TTF)$_2$Cu(NCS)$_2$, and $\kappa$-(BEDT-TTF)$_2$I$_3$ at ambient pressure, respectively. The
phase transition between the (antiferromagnetic) Mott insulator and the
metal/superconductor can be explored by gradually replacing Cl by Br in
\etbrcl. Here
$c$ and $a$ are the lattice parameters;
$b1$ and $p$ indicate the transfer integral
according to Fig.~\protect\ref{fig:kappapattern1}a.
$U/t$ is the on-site Coulomb repulsion with respect
to the hopping integral $t$.}
\end{figure}

In the $\kappa$-phase crystals, conducting layers of cationic
bis-(ethyl\-ene\-di\-thio)\-te\-tra\-thia\-ful\-va\-lene
(BEDT-TTF) molecules are separated by `charge-reservoir' layers of
monovalent anions. As depicted in Fig.~\ref{fig:kappapattern1},
two BEDT-TTF$^{+0.5}$ molecules form confacial dimers which can be
considered as lattice sites; due to this dimerization the
conduction band is half filled. The anion size sensitively
influences the physical properties of the system very similar to
the variation of pressure, since they define the spacing between
the molecules (molecular sites) and thus the width of the
band.\cite{Mori99,remark4} The ratio of electronic correlations to
the width of the conductance band is the control
parameter\cite{McKenzie97} in the phase diagram depicted in
Fig.~\ref{fig:ETphasediagram}. The ground state  of
$\kappa$-(BEDT-TTF)$_2X$ salts can be switched between an
antiferromagnetic insulating, a superconducting, and a metallic
state. These salts exhibit the highest superconducting transition
temperature of all organic superconductors with $T_c =
12.5$~K.\cite{Jerome94,Ishiguro98}

At ambient temperature the studied $\kappa$-phase BEDT-TTF salts
have common properties, which may be characterized as a narrow-gap
semiconductor or ``bad metal''. When the temperature drops below a
so-called coherence temperature $T_{\rm coh}\approx 50$~K on the
right side of the phase diagram, the metallic behavior becomes
dominant due to the formation of Fermi liquid
quasiparticles\cite{Georges96,Merino00a} until a second-order
transition occurs to a superconducting state. The nature of
superconductivity in organic crystals is subject to discussion for
twenty years\cite{Lang03} but in the present study we focus on the
metallic and insulating states. On the left side of the phase
diagram (Fig.~\ref{fig:ETphasediagram}), i.e., for higher values
of $U/t$, the system never shows metallic properties, but is
gradually driven into an insulating state by electronic
correlations as the temperature drops below 90~K; at $T_N\approx
35$~K magnetic order is observed. NMR measurements in deuterated
samples $\kappa$-($d_8$-BE\-DT\--TTF)$_2$\-Cu\-[N\-(CN)$_{2}$]Br
(which fall right on the phase boundary) revealed that at
low temperatures the transition between the commensurate
antiferromagnet and pseudogapped superconductor is of first
order.\cite{MIY02} Most recently, enormous research efforts were
dedicated to the metal-to-insulator transition and critical
end-point in this highly correlated two-dimensional electron
system. The critical behavior in the vicinity of the Mott
transition was investigated by dc measurements under external
pressure and in magnetic field.\cite{Lefebvre00,LIM03,KAG04} The
alloyed series \etbrcl, studied in the present work, covers the
most interesting region of the phase diagram spanned by the pure
Cl and Br salts, including the border between the Mott insulating
and metallic phases. Our infrared reflection measurements of a
series of compounds with Br concentration $x$ varying between 0
and 90 \% make it possible to explore the temperature and
correlation (bandwidth)-dependent charge dynamics on crossing this
phase boundary, as well as the unusual physical properties in the
metallic region above the superconducting transition.

Several optical experiments were performed on the pristine
compounds \etbr\ and
\etcl\ over the years.\cite{Eldridge91b,Tamura91,Kornelsen92,Eldridge96b,VLA96,McGuire01,SAS04}
They gave a general idea of the electronic excitation observed in
the infrared region:  a broad mid-infrared band around $2500 -
3500$~\cm\ and a narrow Drude-like peak in the spectra of
superconducting Br-compound\cite{remark5} at temperatures below
50~K. In the discussion Section \ref{sec:MIR} we review the
different interpretations of the mid-infrared spectra. Our
investigation of the alloys gives an unambiguous assignment of the
spectral features in this region, important for the analysis of the
charge dynamics in these salts.
BE\-DT\--TTF)$_2$\-Cu\-[N\-(CN)$_{2}$]\-Br$_{0.5}$Cl$_{0.5}$ is
the only mixed compound which has previously been investigated by infrared
spectroscopy,\cite{VLA93,DRO94,VLA96,Pet02,Troung97} but only in the mid-infrared range.

Here, we present for the first time a systematic optical study of the series
\etbrcl, with $x$ crossing all relevant regions of the phase
diagram from the insulating/antiferromagnetic to the
metallic/superconducting state. Our experiments cover a broad
spectral range from 50 to 10000~\cm\ and temperatures from room
temperature down to $T=5$~K. This enables us to follow the
response of the free and localized carriers for the different
points of this phase diagram, depending on temperature and
correlation-to-bandwidth ratio. While we focus on the signature of
localized charge excitations here, the succeeding
paper\cite{Dressel05} (which we refer to as Part~II in the
following) is devoted to the dynamics of free charge carriers and
the formation of the coherent quasiparticle response.

\section{Experiments}
Single crystals of the  \etbrcl\ salts were grown by standard
electrochemical methods. Certain ratios of Br/Cl concentration were
chosen to obtain a series of alloys. Subsequent to the optical
reflection experiments, each individual crystal was checked by
microprobe analysis in order to determine the composition. The
actual Br/Cl ratio turned out to be significantly different than
expected from the starting concentration, since Br enters the
compounds much easier than Cl.\cite{remark9} It was hardly possible to produce
samples with predefined Br concentration. In the following, we
denote the actual Br content of
 \etbrcl\ where $x = 0\%$, 40\%, 73\%, 85\%, and 90\%; the concentration was found homogeneous for each specimen. The
platelets contained naturally flat ($ac$) surfaces with a typical
size of about $1\times 1$~mm$^2$.\cite{remark8}
 The orientation was determined from the optical
spectra.

\begin{figure}
\centering
\includegraphics[width=65mm]{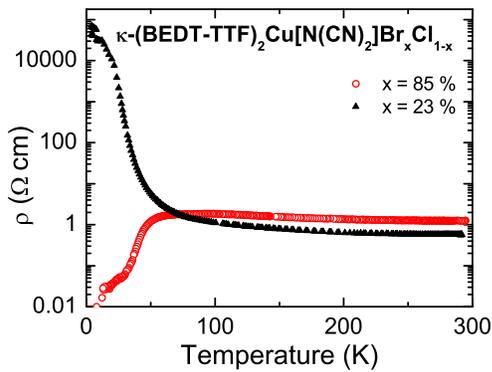}
\caption{\label{fig:dc}Dc resistivity {\em versus} temperature of
$\kappa$-(BEDT-TTF)$_2$\-Cu[N(CN)$_2$]Br$_{0.23}$Cl$_{0.77}$ and
$\kappa$-(BEDT-TTF)$_2$\-Cu[N(CN)$_2$]Br$_{0.85}$Cl$_{0.15}$. While
the latter compound be\-comes metallic below  100~K and eventually superconducting at $T_c=12$~K, the former one gradually turns insulating.}
\end{figure}

The strong dependence of the dc resistivity on the Br/Cl ratio is
emphasized in Fig.~\ref{fig:dc} where $\rho(T)$ is plotted for a
crystal with low ($x=23\%$) and high ($x=85\%$) bromine content. For
the insulating sample
$\kappa$-(BEDT-TTF)$_2$Cu[N(CN)$_2$]Br$_{0.23}$Cl$_{0.77}$ the
resistivity rises by many orders of magnitude as the temperature
decreases, most dramatically below 70~K. Contrary,
$\kappa$-(BEDT-TTF)$_2$Cu[N(CN)$_2$]Br$_{0.85}$Cl$_{0.15}$ shows
basically the same temperature dependence like the pure Br specimen:
$\rho(T)$ increases slightly below room temperature until it reaches
a broad maximum around 100~K. At lower temperatures the behavior is
metallic with $\rho(T)\propto T^2$ for $T < 35$~K and finally a
superconducting transition is observed at $T_c\approx
12$~K.\cite{remark1} The superconducting properties of \etbrcl\ with
different Br concentrations have been investigated previously by dc
resistivity and magnetization.\cite{Sushko93}

The in-plane optical reflectivity was measured with light
polarized along $a$ and $c$ axes, respectively. Employing a
modified Bruker IFS 113v Fourier-transform spectrometer, we
covered a broad frequency range from 50 to 10\,000~\cm\ (6~meV -
1.2~eV) with a resolution of up to 0.5~\cm. The single crystals
were studied at 300, 150, 90, 50, 35, 20 and 5 K with the help of
a cold-finger cryostat. To achieve good thermal contact, the
samples were fixed by carbon paste on a brass cone directly
attached to the cold finger. Absolute values of the reflectivity
are obtained by subsequently evaporating gold onto the sample and
remeasuring it as reference mirror at all
temperatures.\cite{HOM93} The {\it in-situ} gold-evaporation
technique is more accurate than other referencing methods because
it utilizes the entire sample surface and is less effected by
surface imperfections. In addition, for the crystals with 40\% and 85\% Br
concentration, reflectivity spectra were measured in
$2000 - 12\,000$~\cm\ range at temperatures between 300 and 20~K using
a Bruker IFS 66v spectrometer equipped with an IR microscope and cold-finger
Cryovac Microstat. The spectra in this range coincide for 
both methods of measurement; there is basically no dependence on 
Br-content and temperature. 
In the
overlapping range they are in agreement with the room-temperature
data received by Drozdova {\it et al.}\cite{DRO94} up to 40\,000 \cm.
These spectra were also used as a high-frequency extrapolation for the
other compounds. From the reflectivity spectra, the optical
conductivity was calculated employing Kramers-Kronig
analysis.\cite{DresselGruner02}
 At low-frequencies the data were extrapolated
by the Hagen-Rubens behavior, which was double-checked by the dc
resistivity obtained from standard four-probe measurements
(Fig.~\ref{fig:dc}). The low-frequency extrapolation only
very weakly affects the spectra in the measured range, i.e.\ the
absolute values of conductivity.

\section{Results}
\begin{figure}
\centering
\includegraphics[width=85mm]{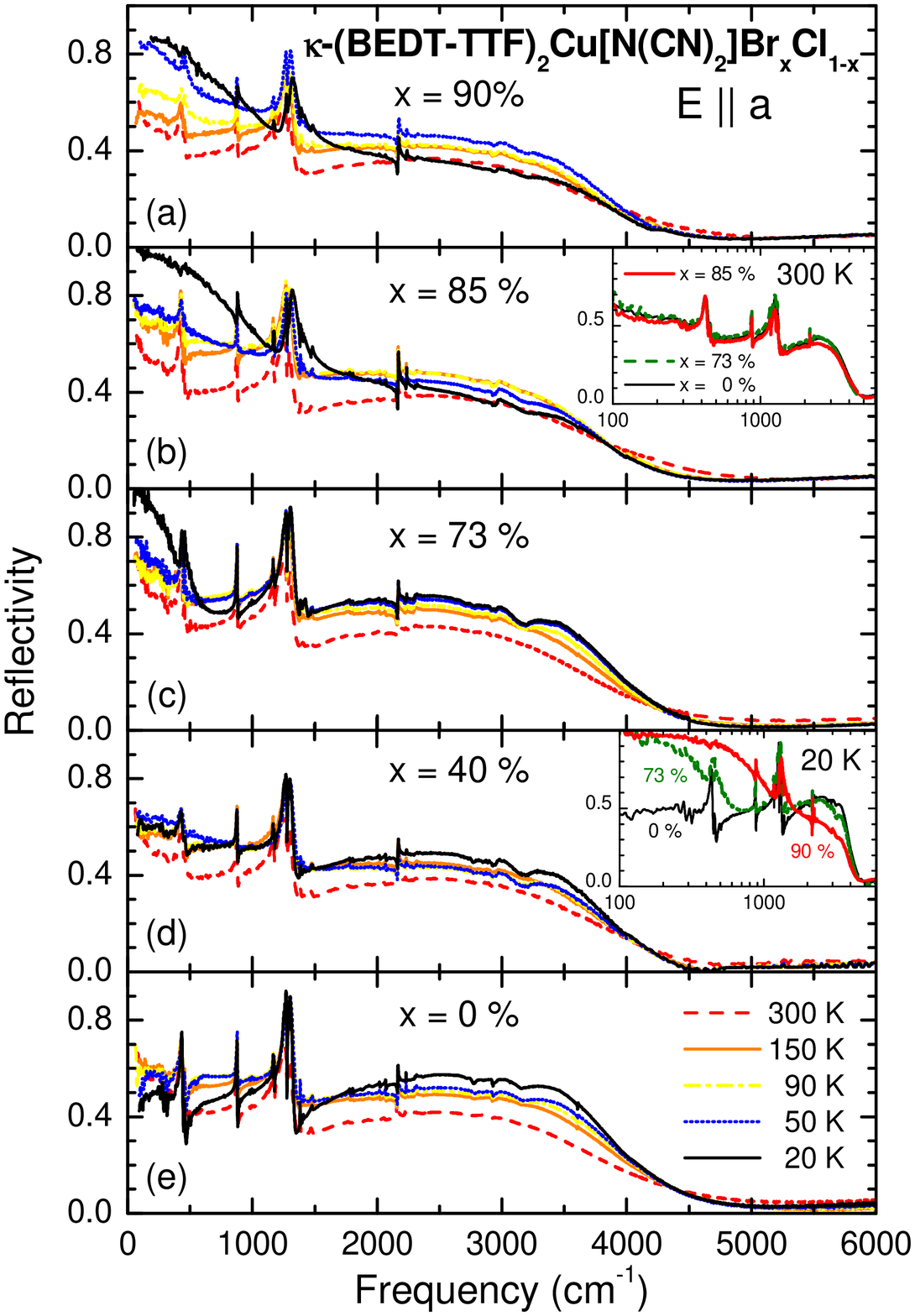}
\caption{\label{fig:ReflA}Reflectivity spectra of  \etbrcl\ for
the polarization $E \parallel a$ measured at various temperatures:
$T=300$~K, 150~K, 90~K, 50~K, and 20~K. The panels (a) - (e)
correspond to different Br concentrations: $x=90\%$, $85\%$, $73\%$,
$40\%$ , and $0\%$. The insets show the room-temperature and
low-temperature spectra for $x=90\%$, 73\%, and 0\%\ on a logarithmic frequency scale.}
\end{figure}

\begin{figure}
\centering
\includegraphics[width=85mm]{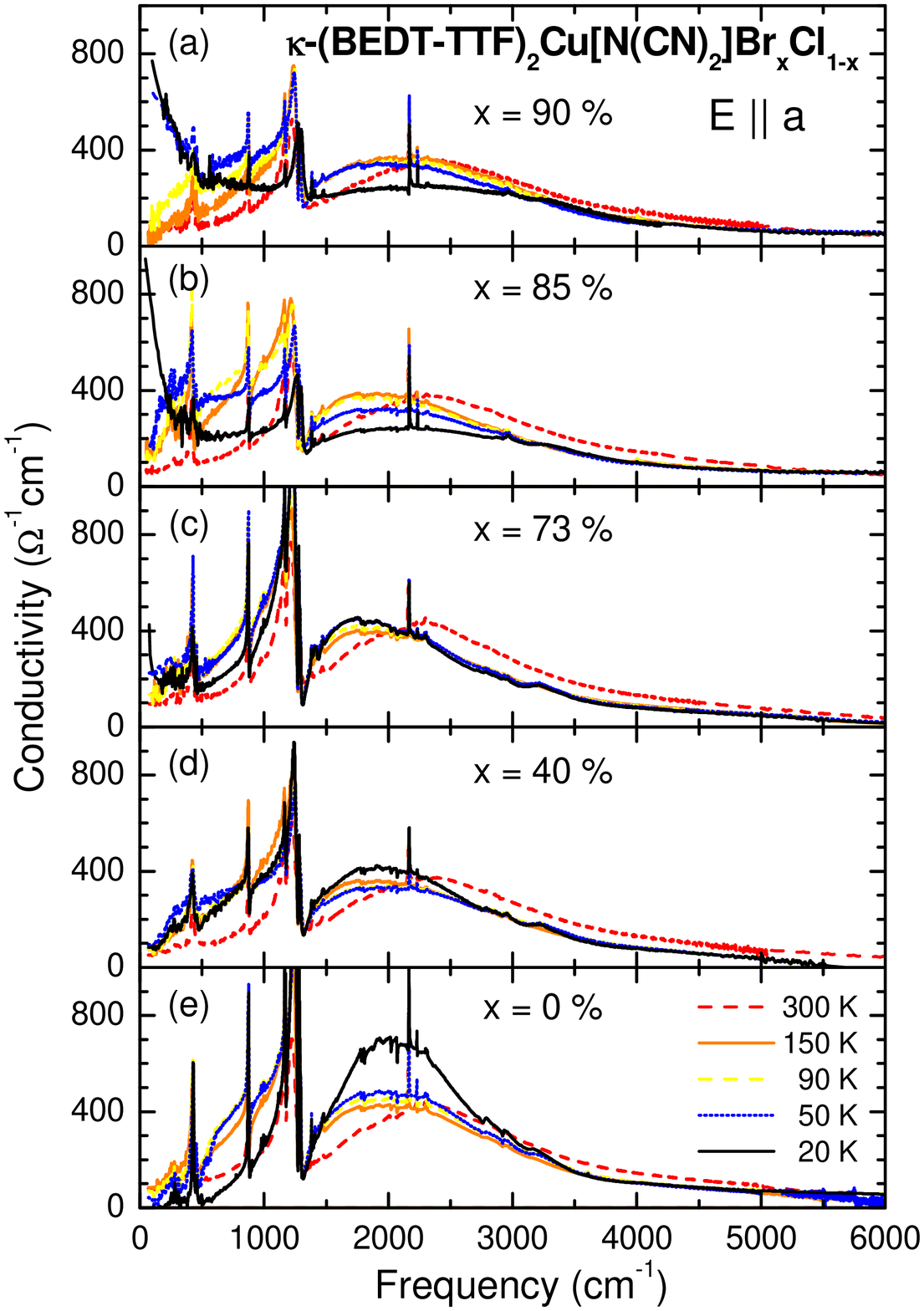}
\caption{\label{fig:LeitA}Optical conductivity spectra ($E
\parallel a$) of  \etbrcl\ at different Br concentrations $x$ and temperatures, obtained by a Kramers-Kronig analysis from the data of Fig.~\protect\ref{fig:ReflA}.}
\end{figure}
In Figs.~\ref{fig:ReflA} and \ref{fig:LeitA} the reflectivity and
conductivity spectra of  \etbrcl\ (with $x=0\%$, 40\%, 73\%, 85\%,
and 90\%)  are plotted for light polarized parallel to the $a$
direction at distinct temperatures from 300~K down to 20~K.
Because there is no significant difference between the $T=20$~K
and 5~K spectra, we omitted the latter. At ambient temperature,
the optical properties only weakly depend on the Br-content (inset
of Fig.~\ref{fig:ReflA}b). As expected for semiconductors, the
reflectivity is basically frequency independent at small
frequencies, and hence the corresponding room-temperature
conductivity is low; the reflectivity starts to decrease
significantly above 3500~\cm\ and reaches a value close to zero in
both polarizations around 5000~\cm. The respective conductivity
spectra show  a broad absorption band centered between 2000~\cm\
and 3000~\cm\ which is well documented in literature for the
$\kappa$-phase of BEDT-TTF salts in general. The strong absorption
features observed in the mid-infrared around 400, 850, and
1400~\cm\ are totally symmetric vibrations of the BEDT-TTF
molecule, activated by electron-molecular vibrational (emv)
coupling (cf.\ Ref.~\onlinecite{Dressel04} and references
therein), we will give a detailed analysis in Sec.~\ref{sec:vibrations}.
\begin{figure}
\centering
\includegraphics[width=85mm]{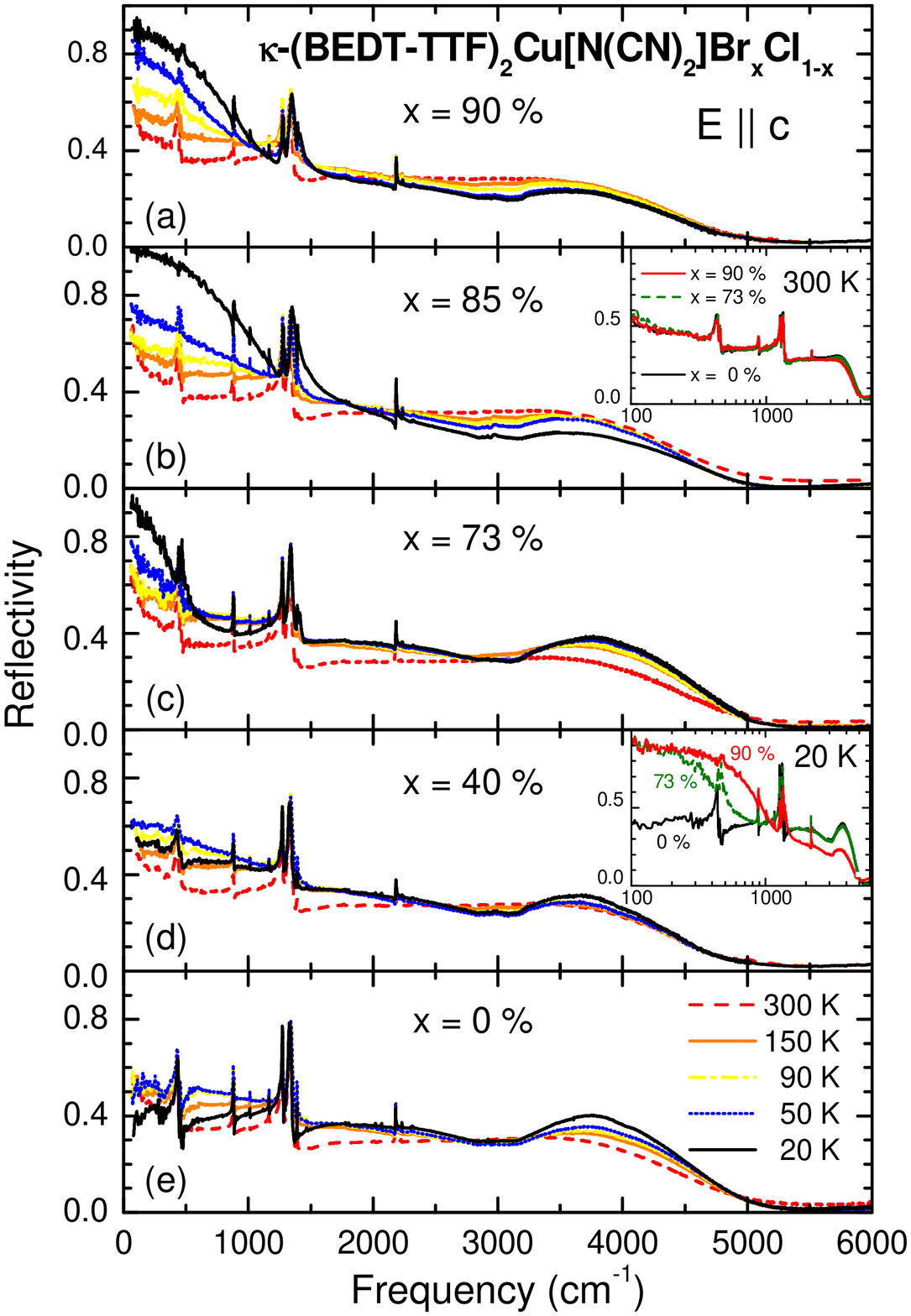}
\caption{\label{fig:ReflC}Reflectivity spectra of  \etbrcl\ for
light polarized $E \parallel c$ measured at various temperatures
as indicated. The different panels (a) - (e) correspond to Br
concentrations: $x=90\%$, $85\%$, $73\%$, $40\%$ , and $0\%$. The
insets show the room-temperature and low-temperature spectra on a
logarithmic frequency scale.}
\end{figure}

Significant changes of the optical spectra are observed when
cooling the samples below $T=90$~K. The far-infrared reflectivity
strongly increases for specimens with high Br content
(Fig.~\ref{fig:ReflA}), while the mid-infrared reflectivity is
suppressed. Correspondingly, as seen in Fig.~\ref{fig:LeitA}, a
Drude-like contribution develops in the conductivity spectra of
the samples with $x=73\%$, 85\%, and 90\%\ at low temperatures.
The latter compound exhibits a behavior similar to the pristine
\etbr\ salt.\cite{Eldridge91b} The opposite is observed  for the
salts with low Br-content: the far-infrared reflectivity drops
while it rises in the mid-infrared. The absolute values of
reflectivity and conductivity are slightly enhanced compared to
previously published results.
\cite{Eldridge91b,Kornelsen92,Eldridge96b,VLA96,McGuire01,SAS04}
This we attribute to our advanced {\it in-situ} gold-evaporation
method for the reference measurement which also accounts for
imperfections of the crystal surface.

\begin{figure}
\centering
\includegraphics[width=85mm]{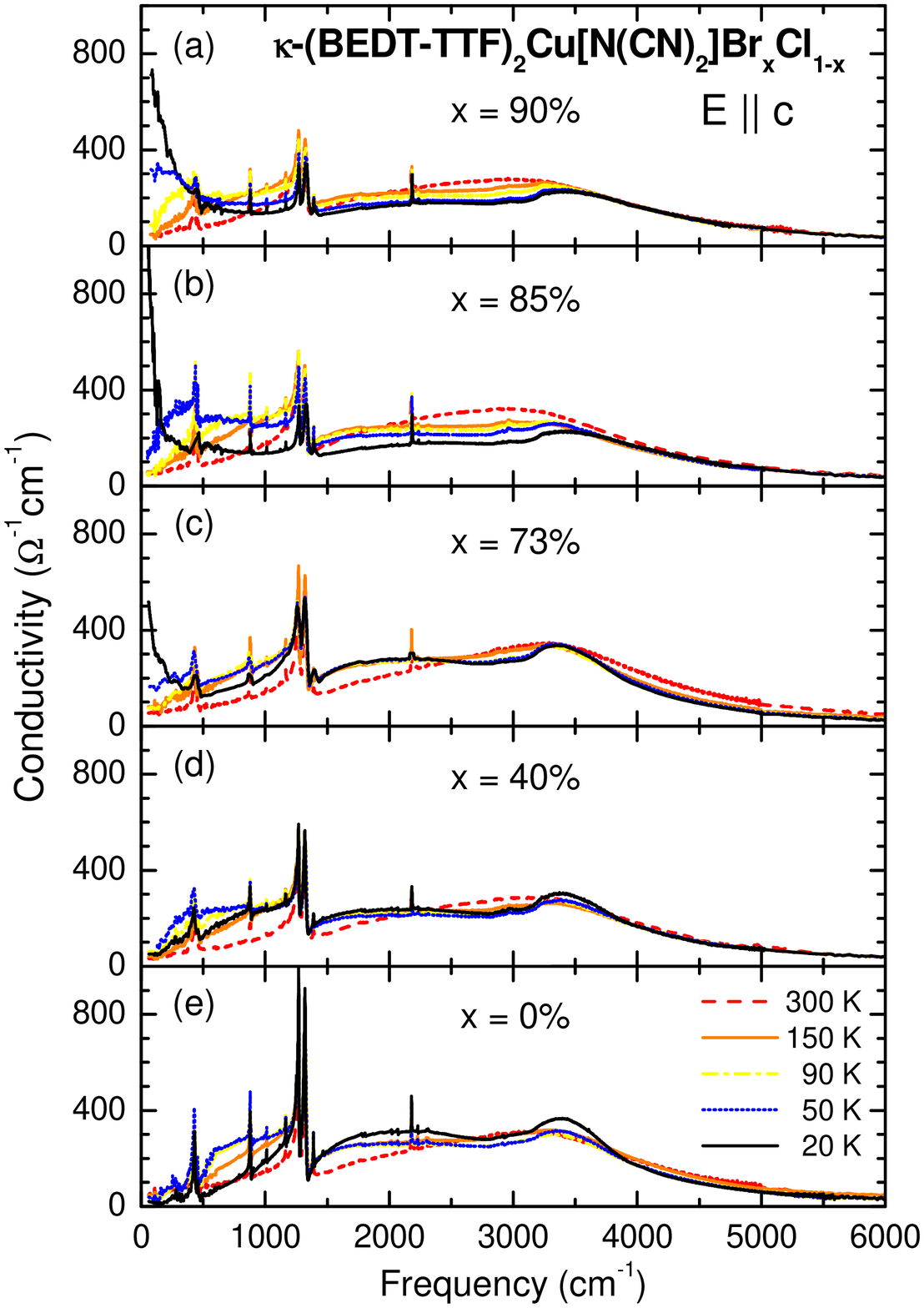}
\caption{\label{fig:LeitC}
Optical conductivity spectra ($E \parallel c$) for  \etbrcl\ of
different Br concentrations (a) $x=90\%$, (b) $85\%$, (c) $73\%$, (d)
$40\%$ , and (e) $0\%$ measured at various temperatures: $T=300$~K,
150~K, 90~K, 50~K, and 20~K. The mid-infrared band clearly has two contributions.}
\end{figure}
The  reflectivity and conductivity for the perpendicular polarization
($E \parallel c$) are shown in Figs.~\ref{fig:ReflC} and
\ref{fig:LeitC} for different Br concentrations $x$ and temperatures
$T$. The spectra exhibit basically the same features as the ones
recorded along $a$ direction; except the shape of the mid-infrared
absorption is different. As previously reported, for most other
$\kappa$-salts, the maximum of the absorption band for the $c$ axis
lies at higher frequencies. While at ambient temperature a distinction
is difficult, at low temperatures it becomes obvious from both,
reflectivity and conductivity data, that it consists of two components:
in addition to the band around 2000~\cm, a second narrower mode has its
maximum  around 3500~\cm. This behavior is most pronounced for the pure
Cl compound. Again, with increasing Br content a Drude contribution
develops as the temperature is reduced below 50 K.

We want to point out that the accessible frequency range of our
experiments is limited to $\nu \ge 50$~\cm\ due to the small sample
surfaces.  Thus, we cannot detect the superconducting energy gap
$\Delta_0$, which is expected around $2\Delta_0=3.53 k_BT_c\approx
30$~\cm.

\section{Discussion}
\label{sec:discussion}
Despite the above mentioned limitations, our
data cover a very broad frequency range from 50 to 10\,000~\cm.
Therefore, we are able not only to study the vibrational features
and the mid-infrared absorption, but also to analyze the temperature
and doping dependence of the Drude contribution in \etbrcl.
However, since there is no agreement in the literature on the
interpretation of the mid-infrared absorption band which is
important for analysis of the whole charge dynamics in these
materials, we first address this part of the spectra. The
contribution of the itinerant electrons will extensively be analyzed and
discussed in Part~II.

\begin{figure}
\centering
\includegraphics[width=75mm]{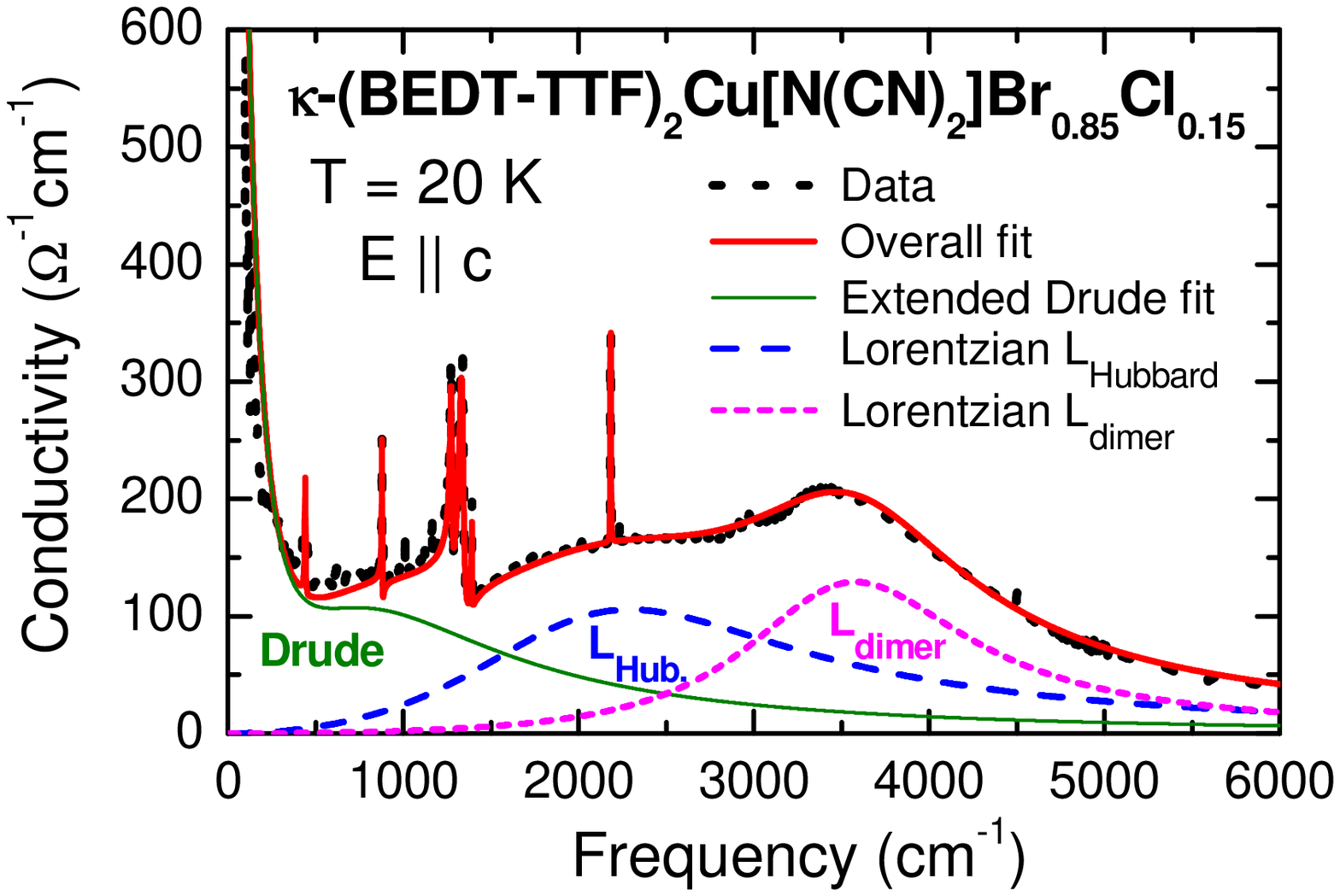}
\caption{\label{fig:Oszillator} Fit of the frequency dependent
conductivity of
$\kappa$-(BE\-DT\--TTF)$_2$\-Cu\-[N\-(CN)$_{2}$]\-Br$_{0.85}$Cl$_{0.15}$
at 20~K for the electric field polarized parallel to the $c$ axis.
L$_{\rm dimer}$ and L$_{\rm Hubbard}$ show the two Lorentz
oscillators required to fit the mid-infrared peak. The narrow lines
describe the vibrational features.
In addition an extended Drude model is
included for the low-frequency increase.}
\end{figure}
The common approach is to fit the spectra  by the Drude-Lorentz
model,\cite{DresselGruner02,Dressel04} because it helps to
disentangle the contributions of conduction electrons, interband
transitions and vibrational features. As an example, the  optical conductivity
along the $c$-direction of the crystal with $x=0.85$ is displayed in
Fig.~\ref{fig:Oszillator} together with a fit by one Drude-like component
and several Lorentzian oscillators; to further restrict the
parameters, the R($\omega$) and $\sigma(\omega)$ spectra were fitted
simultaneously. This procedure was applied to the spectra taken in
both polarization directions and at all temperatures and values of
Br doping.

\subsection{Mid-Infrared band: overview of the different interpretations}
\label{sec:MIR} 
In general, the most prominent feature in the
optical conductivity spectra of $\kappa$-phase BEDT-TTF salts is the
broad mid-infrared hump. It  peaks around 2200~\cm\ for the
polarization $E\parallel a$ and at 3200~\cm\ in the $c$ direction,
where it exhibits a more complicated double structure. Albeit it
seems obvious -- in particular when cooling down -- that two
contributions add up for this band, the explanations proposed over
the years took into account only one single process. It was
suggested that the mid-infrared peak is due to charge transfer
inside the
dimer\cite{Sugano89,Eldridge91b,Tamura91,Kornelsen92,Kornelsen92b}
or due to transitions between the Hubbard bands formed by the
correlated conduction
electrons.\cite{Rozenberg95,McKenzie98,SAS04}

Eldridge and coworkers \cite{Eldridge91b,Kornelsen92} first
suggested that the mid-infrared peak is due to charge-transfer
bands with the excitations confined to the dimers and the charge
transfer occurring between adjacent molecules. The polarization
dependence is explained by different interactions between
neighboring dimers (Fig.~\ref{fig:kappapattern1}a).
The arguments were supported by electronic
band-structure calculations of the $\kappa$-phase salts performed
by Whangbo and collaborators on the basis of the tight-binding
approximation.\cite{Jung89,Geiser91} The highest occupied band is
half filled with only very little difference in bandwidth and
density of states at the Fermi level when going from \etcl\ to
\etbr. Obviously this is a very rough approximation which can
explain neither the semiconducting behavior at ambient temperature,
nor a redistribution of the spectral weight from this mid-infrared
maximum to the Drude-peak for compounds with high Br concentration
on cooling (discussed in Part II), nor the different ground
states of the compounds.

A more consistent picture was achieved when both components of the
optical conductivity, the Drude-like contribution and  the
mid-infrared hump were explained in the framework of a half-filled
two-dimensional system with strong electronic
correlations.\cite{Kino96,Mori99,Seo04} The structure of the
$\kappa$ phase was mapped on an  anisotropic triangular lattice,
each site presenting one dimer as depicted in
Fig.~\ref{fig:kappapattern1}b. The interdimer overlap integrals
define the hopping $t$ between the sites of a triangular
half-filled lattice considered by theory, they are $t_1=30$~meV
and $t_2=50$~meV.\cite{Kino96,Seo04}  The intradimer overlap plays
the role of the effective on-site Coulomb interaction
\Ueff.\cite{Kino96,Kanoda97,Mori99} Consequently, \Ueff\ increases
with dimerization: for the Cl-analog b1 is slightly higher which
causes larger \Ueff. {\it Ab-initio} calculations by Fortunelli
and Painelli\cite{Fortunelli97} give the value of \Ueff$=0.4$~eV
for a dimer in \etbr; this is in agreement with experiments.

It has been predicted by theory that the Mott-insulator transition
in a two-dimensional lattice typically occurs when $U$ is
comparable to the bandwidth $W=8t$. In the present case $U_{\rm
eff}/t\approx 8$, implying that we are very close to the
metal-insulator transition.
 \Ueff$/t$
increases when going from Br to Cl anions, i.e.\ moving from right
to left in the phase diagram (Fig.~\ref{fig:ETphasediagram}) and
leads to localization of the charge carriers.\cite{remark2}
Following these considerations, recently\cite{SAS04}  the
experimentally observed mid-infrared band around 3000~\cm\ was
associated with the transition between the Hubbard bands at
$\hbar\omega \approx$ \Ueff.

The application of dynamical mean-field theory to the metallic
side of the phase diagram suggests\cite{McKenzie98,Merino00a} that
besides the mid-infrared band around \Ueff, a quasiparticle peak
at the Fermi level grows with temperature below $T_{\rm
coh}\approx 0.1t^*$, where $t^*$ is the overlap integral.  Due to
transitions between the coherent quasiparticle band and the
Hubbard bands, a new peak is supposed to develop around
$\hbar\omega\approx$ \Ueff$/2$ for $T<T_{\rm coh}$.

\subsection{Mid-Infrared band: our experimental results}
Our present investigation of the substitution series  \etbrcl\
sheds new light on the above formulated controversy because all
features were traced when going from the metallic to the
insulating phase both by changing temperature $T$ and the relative
correlation strength $U/t$. This enabled us to disentangle the
components coming from charge transfer inside the (BEDT-TTF)$_2$
dimers and between the dimers. We propose that the intradimer
transitions cause the high-frequency contribution L$_{\rm dimer}$,
while the band L$_{\rm Hubbard}$ at lower frequencies is ascribed
to interdimer transitions, i.e., to the transitions between the
Hubbard bands.

Generally, the band of the intradimer transition is expected to
appear at higher frequencies compared  to the transition between
the dimers. It was already shown for molecular chains that the
high-frequency band vanishes when the dimerization is
reduced;\cite{Meneghetti91} calculations were also performed
on a two-dimensional $\kappa$-like structure leading to similar
conclusions.\cite{Visentini98} The same result is obtained by the
cluster model,\cite{YAR96} when extending it to tetramers and
hexamers; the electronic transitions summing up the intradimer and
interdimer excitations shift to lower frequencies and get broader
compared to the pure intradimer one.

In both reflectivity and conductivity spectra
(Figs.~\ref{fig:ReflC} - \ref{fig:Oszillator}) for light polarized
parallel to the $c$ axis, two bands can be clearly distinguished:
a narrow high-frequency peak L$_{\rm dimer}$ and a broad peak
$L_{\rm Hubbard}$ located at lower frequencies. Contrary, in
$a$-direction there is no clear separation visible between L$_{\rm
dimer}$ and $L_{\rm Hubbard}$ in the optical spectra
(Figs.~\ref{fig:ReflA} and \ref{fig:LeitA}). Such an anisotropy in
the mid-infrared range is well documented for these
$\kappa$-salts.\cite{Dressel04,Drichko04} However, the detailed
analysis based on the cluster model given below will show that
despite the anisotropy both peaks are present in either
orientations.
\begin{figure}
\centering
\includegraphics[width=85mm]{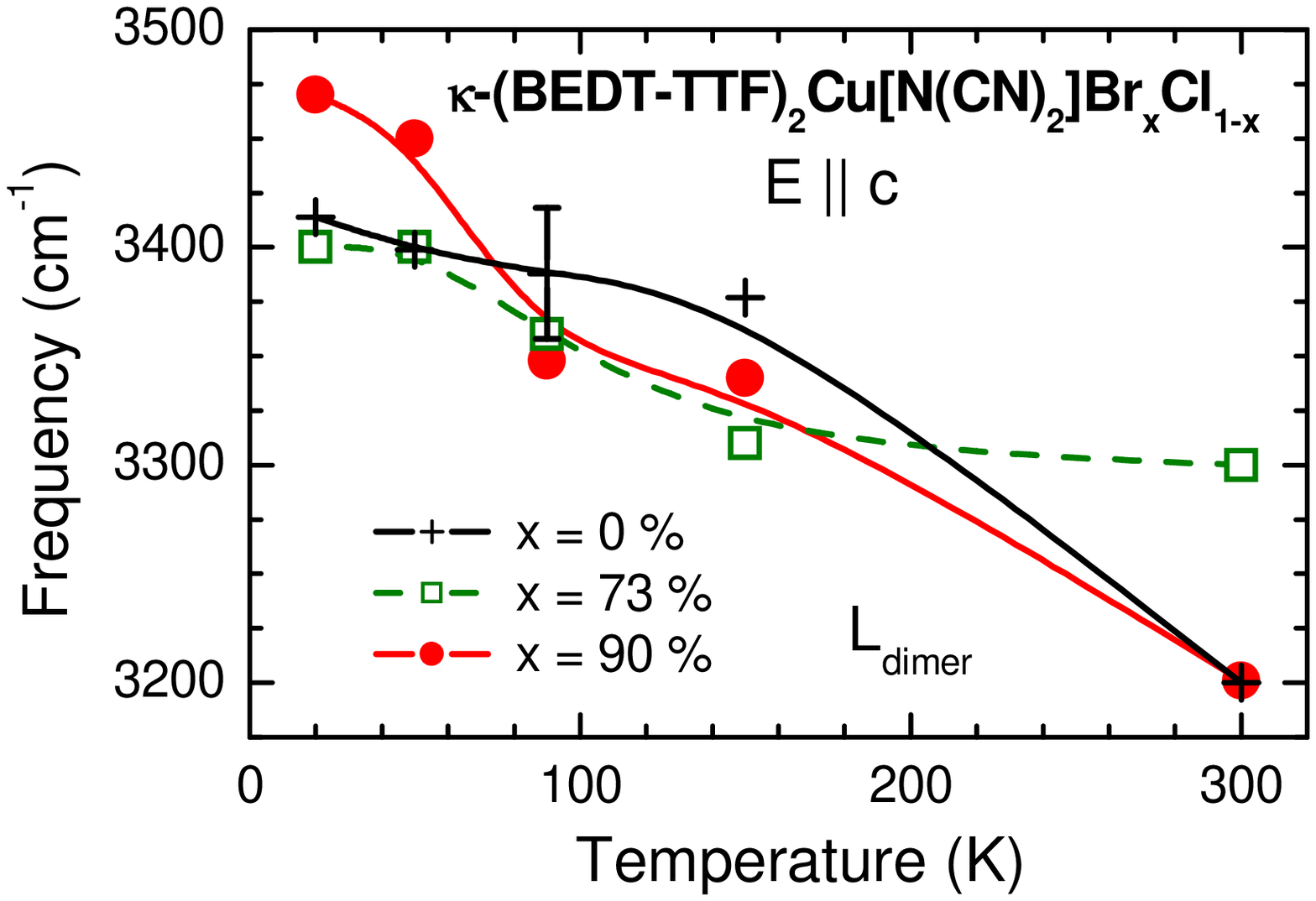}
\caption{\label{fig:I10} Temperature dependence of the
mid-infrared peak L$_{\rm dimer}$  in  \etbrcl\ with $x=0\%$,
73\%, and 90\%\ obtained by a Lorentzian fit for the polarization
$E
\parallel c$. The lines are guides to the eye.}
\end{figure}

The temperature dependence of the L$_{\rm dimer}$ frequency
($E\parallel c$) is plotted in Fig.~\ref{fig:I10} for different Br
concentrations. Within the uncertainty of the Lorentz fit, at
ambient temperature the position of the high-frequency oscillator
L$_{\rm dimer}$ does not show a distinct dependence on the Br
content along the $c$ axis; also the oscillator strength of this band
does not vary substantially. In general there is an upward shift
of approximately 100 to 200~\cm\ when going down to $T=20$~K. We
attribute this blue shift to the thermal contraction of the single
crystals which slightly enhances the intradimer transfer integral
b1 (Fig.~\ref{fig:kappapattern1}a), i.e. the intradimer overlap
increases\cite{Mori99}. The difference between metallic and
insulting compounds is seen at 50~K and lower: while the
temperature dependent high-frequency shift of L$_{\rm dimer}$ is
more enhanced upon cooling in the metallic samples with $x=90\%$
and 85\%, it levels off for the insulating ones with low Br
concentration. Obviously, the L$_{\rm dimer}$ oscillator frequency
is not considerably affected by the opening of the Mott-Hubbard
gap at $T\leq 50$~K which results in an increase of the dc
resistivity of several orders of magnitude.

The intradimer transition, i.e., a charge transfer between the two
face-to-face arranged BEDT-TTF molecules is expected to be strongly
coupled to the totally symmetric molecular vibrations. Thus an
detailed analysis of the experimentally obtained temperature
dependence of their spectra and a comparison to the theoretical
predictions of the cluster model will be the key to a final
assignment of the electronic features in the spectra.

\subsection{Vibrational features
\label{sec:vibrations}}
The sharp absorption features in the frequency range between
400~\cm\ and 1600~\cm\ are known to be totally-symmetric vibrations
activated by coupling with electronic excitations.  In the course of
numerous vibrational studies on the $\kappa$-phase BEDT-TTF salts,
Eldridge's
group\cite{Kornelsen92b,Eldridge95,Eldridge96b,Eldridge97} and
others\cite{DRO94,McGuire01} presented a complete assignment of the
totally-symmetric vibrations. Here we use the C$_{2h}$ symmetry
assignment which takes a deformation of a BEDT-TTF molecule inside
the crystal into account.\cite{remark3} In our work we could follow
not only the temperature, but also the doping-dependence of these
features, that improves our interpretation of the spectra.

\begin{figure}
\centering
\includegraphics[width=85mm]{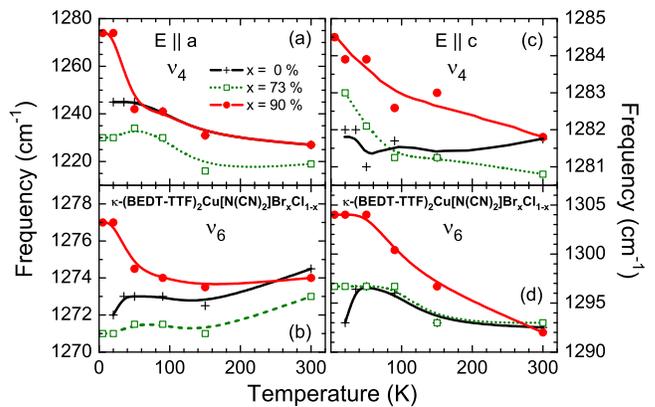}
\caption{\label{fig:nu3}\label{fig:nu5} Temperature dependence of
the mode frequencies of the intramolecular vibrations  $\nu_4({\rm
A}_g)$ and $\nu_6({\rm A}_g)$ for  \etbrcl with Br concentrations
0 \%, 73 \%, and 90 \%. (a)~The $\nu_4({\rm A}_g)$ mode along the
$a$-direction was fitted by a Lorentzian and the center frequency
plotted. (b)~In the case of $\nu_6({\rm A}_g)$ the frequencies are
defined as the minima in the respective optical conductivity,
i.e.\ the strongest of the quadruple. In frames (c) and (d) the
equivalent data are presented for the polarization $E\parallel c$.
The lines correspond to spline fits.}
\end{figure}

The most prominent $\nu_4({\rm A}_g)$ vibration involves a
symmetric stretching of the C=C double bonds;\cite{remark6} at
room temperatures it is observed at about 1240~\cm\  along the
$a$-axis and at 1280~\cm\  along the $c$-axis in the spectra of
the  \etbrcl. We estimate the center frequency of the $\nu_4({\rm
A}_g)$ band by fitting it with a Lorentzian, disregarding the
anti-resonances due to the $\nu_6({\rm A}_g)$ mode for a moment. As
can be seen from Fig.~\ref{fig:nu3}a and c, with decreasing
temperature ($300~{\rm K}\leq T < 50$~K) the $\nu_4({\rm A}_g)$
modes slightly shift to higher frequencies in about the same
manner for all compounds. As $T$ is reduced this tendency enhances
further for $x=85\%$ and 90\%: the total shift amounts to
approximately 45~\cm\ for $E\parallel a$, and less than 5~\cm\ in
$c$ direction; the variation saturates at very low temperatures.
For the insulating compounds with low Br content, the $\nu_4({\rm
A}_g)$ mode even reverses its temperature dependence below $T_{\rm
coh}\approx 50$~K and becomes softer. Except some gradual
difference, the behavior is very similar for both orientations.
Below 50 K, the $\nu_4({\rm A}_g)$ mode becomes sharper in the
insulating samples while it broadens and seems to be weaker for
high Br content.
\begin{figure}
\centering
\includegraphics[width=85mm]{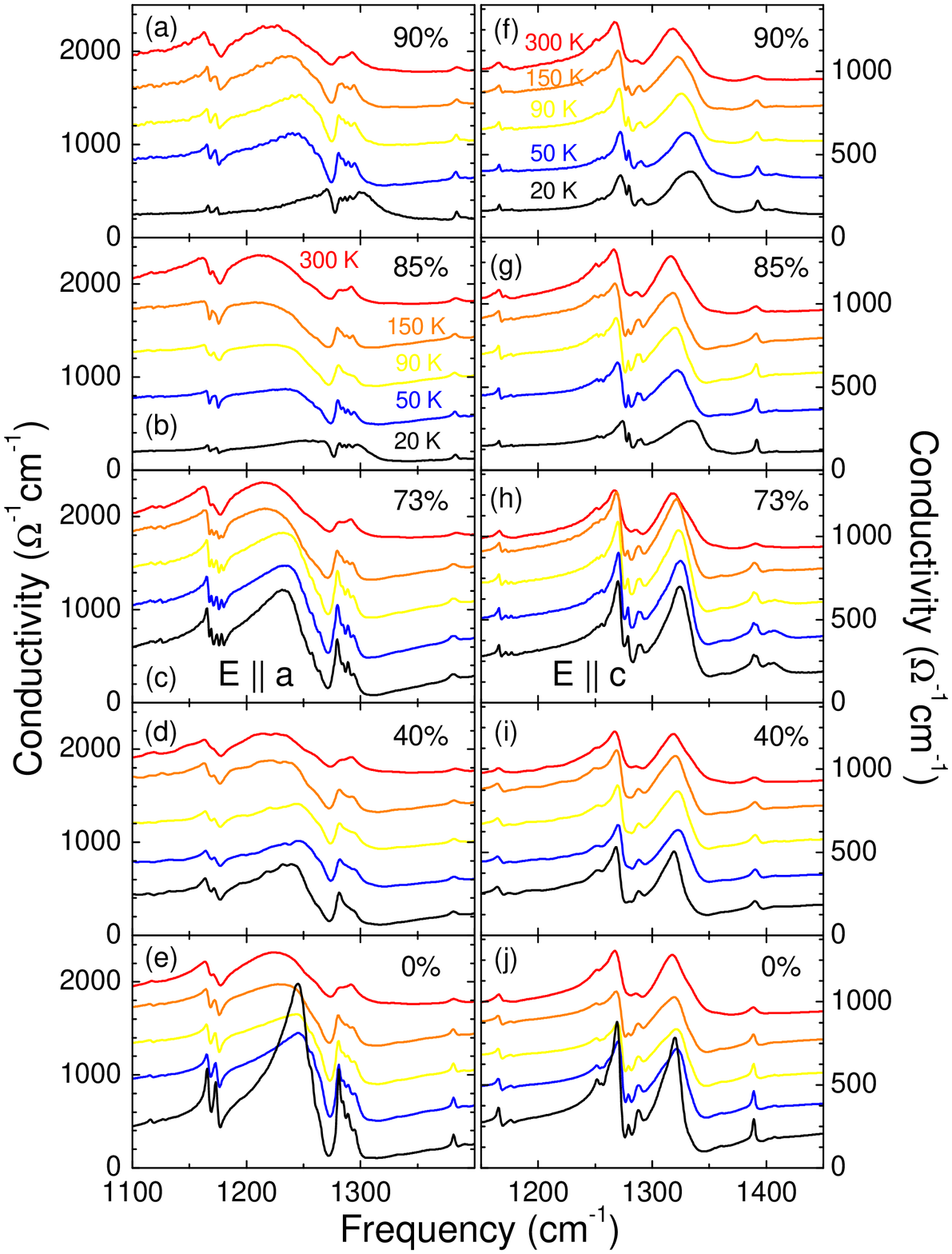}
\caption{\label{fig:phononsa}\label{fig:phononsc} Detailed view of
the optical conductivity of  \etbrcl\ along the $a$ and
$c$-directions (left and right panels, respectively) for different
Br  concentrations $x=90\%$, 85\%, 73\%, 40\%, and 0\%, (a)-(e)
and (f)-(j), respectively. The spectra for the different
temperatures  (bottom to top: $T=20$~K, 50~K, 90~K, 150~K, and
300~K) are offset by 400~($\Omega{\rm cm})^{-1}$ for clarity.}
\end{figure}

The frequency of the $\nu_6({\rm A}_g)$ mode (vibration of the CH$_2$
groups) overlaps with the broad emv-coupled $\nu_4({\rm A}_g)$
band. In Fig.~\ref{fig:phononsa} the excitation is seen as an
antiresonance around 1270 to 1280~\cm\ for $E\parallel a$ and
slightly higher for the perpendicular direction. At low
temperatures, four bands of $\nu_6({\rm A}_g)$ are resolved,
originating from four distinct CH$_2$ groups  per unit cell. We
follow the  temperature dependence by choosing the minimum around
1273~\cm. Again, for low Br content the $\nu_6({\rm A}_g)$ mode
gradually moves down in frequency with decreasing temperature,
while it significantly shifts to higher values for $x=85\%$ and
90\% as presented in Fig.~\ref{fig:nu5}c. The same temperature
dependence of $\nu_6({\rm A}_g)$ is observed parallel to $c$
(Fig.~\ref{fig:nu5}d). A similar temperature and Br-concentration
dependence is seen for the peaks of the emv-coupled
`ring-breathing' mode $\nu_{10}({\rm A}_g)$ of the BEDT-TTF
molecule\cite{Wesolowshi05} at 870~\cm\ and 885~\cm\, and of the
$\nu_{13}({\rm A}_g)$ at about 430~\cm. Accordingly, the peaks are
pretty intense in the insulating state, and are less significant
in the spectra for high Br concentration.

In contrast to the emv-coupled modes, the infrared active
$\nu_{45}({\rm B}_{2u})$ vibration of BEDT-TTF molecule detected
around 1383~\cm\ (Fig.~\ref{fig:phononsa}) and CN-stretch vibration
of the anion layer observed around 2160~\cm\ do not depend on the
Br-concentration and show no pronounced temperature dependence
besides the expected hardening with cooling. Therefore, since the
charge on the BEDT-TTF molecules is not redistributed in \etbrcl
when the samples are cooled down, the characteristic temperature
dependence of the emv-coupled A$_g$ modes has to be due to the
electronic excitation to which they are coupled.

\subsection{Excitations localized on dimers: charge transfer and
emv-coupled features}
The cluster model of M.J. Rice\cite{Rice76} and Delhaes and
Yartsev \cite{Yartsev93,YAR96} describes the optical properties
of molecular clusters with arbitrary geometry and equilibrium
charge density distribution.  The model describes optically
activated charge transfer between the molecules in a cluster, the
parameters defining this transition are transfer integral $t$ 
between the molecules and Coulomb repulsion $U_{\rm mol}$ of two
electrons on one molecule. In addition, this model takes into
account an activation of totally symmetric vibrations by this
charge transfer in a cluster; the strength of this coupling
is defined by the coupling constants $Q_i$, specific for each given
molecular vibration. It gives a correlation between a charge
transfer electronic transition and the emv-coupled features. The
vibrational modes become infrared active by emv coupling to the
charge-transfer excitation; and they are shifted down in frequency
with respect to the corresponding Raman
modes.\cite{Eldridge95,Eldridge96b,Eldridge97} The shift and the
intensity of the emv-coupled features depend on the coupling
constants and on the position of the respective charge-transfer
band.
 The model taking into account two perpendicular dimers of BEDT-TTF molecules successfully describes the
mid-infrared peak and emv-coupled features in the room-temperature
spectra of the $\kappa$-phase salts,\cite{Vlasova92,YAR96} while
it does not account for the metallic behavior; thus it is not able to
mimic the appearance of a Drude-peak in the compounds with high Br content.

\begin{figure}
\includegraphics[width=8cm]{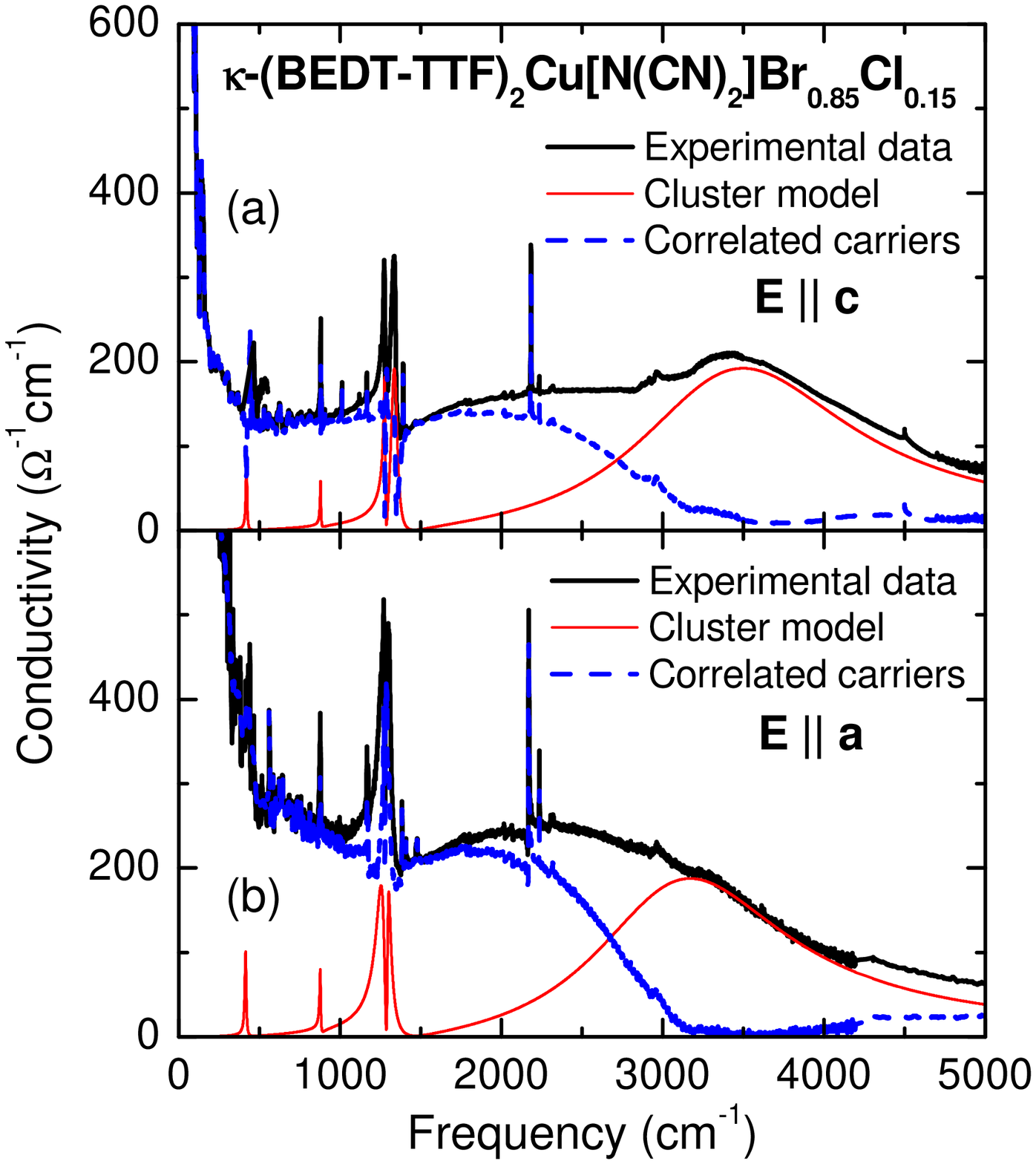}\\
\caption{ \label{Fig:dimermodel}Conductivity of $\kappa$-(BEDT-TTF)$_2$\-Cu[N(CN)$_2$Br$_{0.85}$Cl$_{0.15}$ 
for the two polarizations 
(a) $E\parallel c$ and
(b)  $E\parallel a$ at $T=20$~K. 
The solid red curves represent the conductivity for the intradimer transition and
the emv coupled modes calculated within the cluster model. 
The dotted blue curves show the difference between the experimental data (thick black line)
and the red curve.
Fit parameters for $E\parallel c$ are $U_{\rm mol}=0.5$~eV and $t({\rm b1)} = 0.21$~eV. The
vibrational frequencies $\nu_i$ and respective coupling constants $Q_i$ are: 
$\nu_4= 1460$~\cm\ $Q_4=610$, 
$\nu_6= 1287$~\cm\ $Q_6=95$, 
$\nu_{10}= 878$~\cm\ $Q_{10}=120$, and
$\nu_{13}= 445$~\cm\ $Q_{13}=240$.
For $E\parallel a$ we used $U_{mol}$=0.5 eV and $t({\rm b1}) = 0.19$~eV; the frequencies and coupling constants are the same except 
$Q_4=680$.}
 \end{figure}

In this work we use the simplest dimer model to describe the
emv-coupled features present also in the metallic phase, and to
verify the assignment of the mid-infrared peak L$_{\rm dimer}$ to
the interdimer charge transfer. As an example, in
Fig.~\ref{Fig:dimermodel}a we show a fit by the dimer model 
of the 85 \% Br
compound spectra. For $E\parallel c$ the dimer model with
$U_{\rm mol}=0.5$~eV and transfer integral term $t({\rm b1}) = 0.21$~eV  can
reasonably well predict both the position of the high-frequency
electronic maximum L$_{\rm dimer}$ and of the emv-coupled
features. The value of the transfer integral energy is somewhat lower
than those received by the H{\"u}ckel method:
$t({\rm b1})=0.26-0.27$~eV.\cite{Mori99,Seo04} Taking into account the
very different approaches of these methods, the agreement is quite
good. The differential spectrum of the experimentally obtained conductivity and the
calculations by the dimer model shows that the intradimer
transition is responsible for nearly all the intensity of the
emv-coupled features. With other words, in the low-temperature
metallic state the intradimer transition interacts strongly with
the totally symmetric vibrations, while this process is less
important for the charge transfer between the dimers.

In order to identify the intradimer transition band for
polarization $E\parallel a$, we start with the respective
parameters obtained for the  $E \parallel c$. Only the position of
the intradimer charge transfer is varied  by changing the transfer
integral b1 until the positions of emv-coupled features, which
are considerably softer along $a$ axis, match the experiment.
Consequently, the position of the intradimer transition follows
these emv-coupled vibrations. The fit with $t({\rm b1})=0.19$~eV suggests
that the L$_{\rm dimer}$ has its maximum at about 2900~\cm\ in the
$a$ direction, compared to 3300~\cm\ found for $E\parallel c$.

For a further analysis of the data focused on itinerant charge
carriers, we subtract the results of the cluster-model
calculations from the experimental data.  As seen in
Fig.~\ref{Fig:dimermodel}, the remaining intensity of the
emv-coupled vibrational modes is small; this confirms that they are
only very weakly coupled to charge carriers which are not localized on the
dimers. Interestingly, this is only true for the metallic samples
with high Br content (85\%\ and 90\%). Once we approach the
insulating side of the phase diagram (low Br concentration), a
charge transfer only in a dimer does not account for all the
intensity of the emv-coupled features anymore. These spectra are
much closer to those proposed by the tetramer model,\cite{Vlasova92} 
indicating that once the inter-dimer excitations
get localized they are also coupled to the totally-symmetric
vibrations of BEDT-TTF molecule.


 \subsection{Anisotropy of the spectra}

One of the striking features of the $\kappa$-phase spectra is the
anisotropy, which from the first glance one would not expect on an
orthorhombic unit cell\cite{remark7} with two crystallographically
identical dimers and an angle between the optical axes
and the dimers close to 45 degrees. This anisotropy involves the intradimer transition,
and above we showed that the cluster model gives a good explanation
for the anisotropy in the vibrational features.

It should be noted that the distance between the dimers
is different in $a$ and $c$ direction, with some preference along
the $a$ axis, as depicted in Fig.~\ref{fig:kappapattern1}. In
addition, the molecules are not standing upright on the ($ac$)
plane, but considerably tilted in $a$ direction (in an alternating
fashion in such a way that adjacent layers form a herring-bone
pattern).\cite{Williams92} The higher reflectivity observed in the
E~$\parallel$ a polarization in these compounds suggests that a
projection of the dipole moment for all the electronic transitions
onto this axis is higher.

An explanation for the difference in position and intensity of the
intradimer charge transfer band L$_{\rm dimer}$ observed parallel
to $a$ and $c$ axes might be given by a so-called Davydov
splitting.\cite{Davidov71}. In the present case of the
$\kappa$-salts, the dimers are taken as the principal unit, i.e.\
a `big molecule' with one hole and spin residing on it. In
general, if the unit cell possesses a center of symmetry and two
identical molecules in the cell, a Davydov splitting of the
intra-molecular electronic transitions between the upper molecular
orbitals (intra-molecular excitons, Frenkel excitons) is
observed.\cite{Davidov71} In this case two bands which are
distinct in position and intensity are expected parallel to the
symmetry axes of the unit cell.  Applied to the present case, the
electronic spectrum exhibits two distinct bands for the two
polarizations parallel to the symmetry axes of the unit cell.
Thus, the anisotropy of the intradimer transition can be
considered as a Davydov splitting of the dimer excitation. 
Indeed, the less symmetrical
(monoclinic) $\kappa$-(BEDT-TTF)$_2$Hg(SCN)$_4X$ ($X$=Cl, Br)
 show less anisotropy of the spectra\cite{Vlasova97} than the
presented orthorhombic compounds.

\subsection{Allocating the transition between the Hubbard bands}

From the analysis presented above, we conclude that the
lower-frequency contribution L$_{\rm Hubbard}$ to the mid-infrared
band originates from excitations across the Mott-Hubbard gap which
allows us to determine the effective Coulomb repulsion \Ueff\ to
be approximately  2200~\cm. This assignment is supported by
comparing our data to spectra of the superconductor
$\kappa$-(BEDT-TSeF)$_4$Hg$_{2.89}$Br$_8$. The replacement of the
four inner sulphur atoms with selenium reduces the on-site Coulomb
repulsion but increases the transfer integrals to neighboring
molecules. Therefore, the BEDT-TSeF-based analogs are much closer
to the normal metallic state; the contributions of itinerant and
localized charge carriers are well separated in the optical
spectra. It has clearly been observed\cite{Drichko04} that with
decreasing temperatures the contribution of the electrons in the
conduction band (Drude-peak and transitions between Hubbard bands)
shifts to lower frequencies while there is a blue shift of the
excitations of the localized charge carriers, similar to the
metallic compounds studied in this work.

Calculations by Merino and McKenzie\cite{Merino00b} reveal that in strongly
correlated metals the coupling between electrons and phonons leads to a
non-monotonic temperature dependence of the vibrational modes near the
coherence temperature $T_{\rm coh}$. The shift is most pronounced (up
to 5\%) for phonons in the energy range comparable to \Ueff$/2$, and
becomes weaker for larger or smaller frequencies. Raman measurements
perfectly agree with these predictions.\cite{Lin98} Evidently, the
situation is more complicated for infrared data, as the vibrations
are activated only by emv-coupling and influence in frequency and
intensity by the charge excitations within the dimers. Nevertheless,
this electron-phonon coupling might be an explanation for a softening
of $\nu_4({\rm A}_{g})$ and $\nu_6({\rm A}_{g})$ below $T_{\rm coh}\approx 50$~K
for the compounds \etbrcl\ with $x=0\%$ and 40\%\, where the electronic
correlations $U/t$ are strongest. A close inspection reveals that the
higher frequency mode $\nu_{45}({\rm B}_{2u})$ exhibits a similar
behavior for the insulating samples with low Br content, but weaker;
while no change is observed for the far-infrared vibration
$\nu_{14}({\rm A}_g)$. This also suggests that the effective Coulomb
repulsion \Ueff\ is of the order of 2000 to 2500~\cm.

\section{Conclusion}
Our comprehensive analysis of the temperature- and
Br-concentration dependence of mid-infrared and emv-coupled
features in
$\kappa$-(BEDT-TTF)$_{2}$\-Cu[N(CN)$_{2}$]Br$_{x}$Cl$_{1-x}$ allows
us to distinguish two contributions in the mid-infrared part of
the spectra. This interpretation is supported by the calculations
which describe a charge transfer in a dimer and its coupling to
the totally symmetric vibrations of BEDT-TTF molecule.  The higher
frequency L$_{\rm dimer}$ band (3300~\cm\ for the polarization
$E\parallel c$ and presumably around 2900~\cm\ for $E\parallel a$)
originates from the charge transfer between the BEDT-TTF
molecules in dimers. This charge-transfer within the dimers is
coupled to the intramolecular vibrations of BEDT-TTF and is
responsible for the major part of the emv-coupled features
intensity for the metallic compounds, while in the insulating
materials the lower-frequency contribution is presumably also
coupled with vibrations. The lower-frequency contribution to the
mid-infrared band L$_{\rm Hubbard}$ located at 2200~\cm\ is
isotropic and assigned to transition between two Hubbard bands
which form due to strong electronic correlations. In Part~II
the dynamical properties of the itinerant charge carriers will be analyzed and discussed in detail, including extensive calculations by dynamical mean-field-theory.

\section{Acknowledgments}
The authors are grateful to Alain Barreau for his help at the
microscopic characterization of the sample concentrations. Belal
Salameh performed the dc measurements. We acknowledge the helpful
discussions with Jaime Merino and Ross McKenzie, who initiated the
study many years ago, and A. Girlando, M. Masino, and A. Painelli. 
The project was partially supported by the
Deutsche Forschungsgemeinschaft (DFG). ND is grateful to the
Alexander von Humboldt-Foundation for the continuous support. B.P., V.S. and R.V. thank V. Yartsev for useful
discussions and the algorithm of calculations by a cluster model.


\begin{thebibliography}{99}
\bibitem{remark4}By now, the observations are not completely
understood: the coupling between the layers is certainly crucial,
however, simple volume effects cannot explain the behavior.
\bibitem{remark5}The optical study of an alloy
$\kappa$-(BE\-DT\--TTF)$_2$\-Cu\-[N\-(CN)$_{2}$]\-Br$_{0.5}$Cl$_{0.5}$ was
performed in mid-infrared and could not give an information about
presence or absence of a Drude-peak.
\bibitem{remark8}The small size of the
crystals limited our possibility of low-frequency measurements. In the 
case of the $x=40$\%\ sample, the crystal was so tiny, that
the measurements in the FIR range were only
possible without a polarizer due to the weak signal. 
Since the anisotropy of the spectra
in the FIR range is low, it was possible to use the data for both
polarizations.
\bibitem{remark1}The pure compound \etcl\ follows the rise in $\rho(T)$ down to about 30~K, at lower temperatures the resistivity levels off at a value of approximately $10^3~\Omega$cm and even shows a slight drop around $T=10$~K. This behavior is well documented in literature\cite{Williams90,Ito96,Kobayashi91} and associated with traces of superconductivity induced by internal stress or surface effects. Similar effects are seen in samples with very low Br content $x=15\%$.\cite{Sushko93}
\bibitem{remark3}Previously, the assignment of the modes was done assuming
$D_{2h}$ symmetry of the BEDT-TTF molecule (plane BEDT-TTF
molecule).\cite{Kozlov87,Kornelsen92,Eldridge95,Eldridge96b,Eldridge97,Dressel04}
By now the general agreement is that the molecules in
$\kappa$-(BEDT-TTF)$_2$Cu(NCS)$_2$ only have $D_2$ due to the twisted
ethylene groups \cite{Meneghetti86,Girlando04,Williams92} leading to 19
A$_1$ totally symmetric modes. For the \etbr\ and \etcl-family the
molecules are eclipsed and exhibit only $C_{2h}$ symmetry, as recently
pointed out in Ref.~\protect\onlinecite{Wesolowshi05}.
\bibitem{remark6}As discussed in the following paragraph, the $\nu_4$ mode appears as a double peak due to the $\nu_6$ antiresonance. We fit the complete feature by a Lorentzian  and choose the center frequency, while commonly the upper peak is listed as the $\nu_4$ frequency.
\bibitem{remark7}The dihedral angle is about 92$^\circ$ which makes
the charge transfer between the dimer contribute stronger in the $c$
direction.
\end{thebibliography}
\end{document}